\documentclass[
	aps, aps,pra,longbibliography, superscriptaddress, twocolumn,
	10pt
	floatfix, 
    nofootinbib,
	tightenlines
]{revtex4-1}
\usepackage[final]{graphicx}
\usepackage{times,bbm,amsmath,amssymb}
\usepackage{epsfig,color}
\usepackage{xcolor}
\usepackage{hyperref}
\hypersetup{
    colorlinks = true
}
\usepackage{cleveref}
\usepackage{microtype}

\usepackage{float,siunitx}
\usepackage[caption = false]{subfig}

\usepackage[greek,english]{babel}
\usepackage{thumbpdf,enumerate}
\usepackage{booktabs}
\usepackage{sidecap}
\usepackage[scaled=.8]{couriers}
\usepackage{multirow}
\usepackage{placeins}
\usepackage{relsize}
\usepackage{pst-grad,bm}
\usepackage{epigraph}
\usepackage{gensymb}
\usepackage{longtable}
\usepackage{ulem} 
\usepackage{acronym}
\usepackage{physics}
\usepackage{easyReview}

\usepackage[columnwise]{lineno}

\DeclareUnicodeCharacter{0301}{\'{e}}

\begin{document}


\vspace{10pt}

\title{Single-Image Entanglement Verification with Spatially Encoded Measurement Contexts}

\author{Nazanin Dehghan}
\address{Nexus for Quantum Technologies, University of Ottawa, Ottawa, K1N 6N5, ON, Canada}
\affiliation{National Research Council of Canada, 100 Sussex Drive, K1A 0R6, Ottawa, ON, Canada}

\author{Alessio D'Errico} 
\email{aderrico@uottawa.ca}
\address{Nexus for Quantum Technologies, University of Ottawa, Ottawa, K1N 6N5, ON, Canada}
\affiliation{National Research Council of Canada, 100 Sussex Drive, K1A 0R6, Ottawa, ON, Canada}

\author{Yingwen Zhang} 
\address{Nexus for Quantum Technologies, University of Ottawa, Ottawa, K1N 6N5, ON, Canada}
\affiliation{National Research Council of Canada, 100 Sussex Drive, K1A 0R6, Ottawa, ON, Canada}

\author{Hugo Defienne} 
\address{Sorbonne Universit\'e, CNRS, Institut des NanoSciences de Paris, INSP, F-75005 Paris, France}

\author{Daniele Faccio} 
\address{School of Physics \& Astronomy, University of Glasgow, Glasgow, UK, G12 8QQ}

\author{Ebrahim Karimi}
\address{Nexus for Quantum Technologies, University of Ottawa, Ottawa, K1N 6N5, ON, Canada}
\affiliation{National Research Council of Canada, 100 Sussex Drive, K1A 0R6, Ottawa, ON, Canada}
\affiliation{Institute for Quantum Studies, Chapman University, Orange, California 92866, USA}

\begin{abstract}
Entangled photon pairs produced by spontaneous parametric down-conversion exhibit rich spatial entanglement structure that is often difficult to probe with conventional measurements. Here, we show that spin-orbit optical elements can convert this spatial structure into directly observable quantum interference patterns. Using a $q$-plate, we demonstrate that the relative wavefront curvature of biphoton states generated by a pair of nonlinear crystals can be retrieved from the spatial modulation of coincidence images. Building on this principle, we introduce a liquid-crystal metasurface that performs spatially multiplexed Bell measurements across the transverse profile of the photon field. The device, which we call a Clauser-Horne-Shimony-Holt (CHSH) plate, assigns different polarization projections to different azimuthal sectors of the beam, allowing the sixteen joint measurements required for a CHSH test to be realized simultaneously in a single acquisition. In this architecture, the spatial coordinate acts as a classical register selecting the measurement context, while photon pairs sample these contexts according to their emission directions. We further demonstrate that the same measurement concept can be implemented using a programmable spatial light modulator, providing a dynamically reconfigurable realization of the scheme. Our results show that spatially structured optical elements can transform Bell tests into parallel measurements distributed across the transverse plane, enabling rapid characterization of spatially varying entanglement. This approach opens new possibilities for structured-light quantum measurements, Bell-inequality-based imaging, and the study of spatially engineered entangled photon sources.
\end{abstract}

\maketitle 
\section{Introduction}
Entangled photons have long been recognized as a central resource for a wide range of quantum technologies. They enable higher precision and accuracy in quantum sensing and imaging \cite{giovannetti2011advances,moreau2019imagingwith, couteau2023applications,defienne2024advances, salari2025quantum, derrico2026}, deployment of device-independent quantum communication \cite{pirandola2020advances,yin2020entanglement,basso2021quantum,zhang2022device}, and provide a pivotal tool for quantum information processing protocols, including entanglement swapping~\cite{pan1998experimental} and quantum teleportation \cite{bouwmeester1997experimental,kim2001quantum,hu2023progress}. Among the methods for generating entangled states, spontaneous parametric down-conversion (SPDC) remains one of the most widely used techniques. In this optical process, a high-power pump laser passing through a nonlinear crystal produces, with small probability, a pair of photons which can be entangled in frequency, time, spatial modes and polarization~\cite{ou2007multi,walborn2010spatial}. In particular, polarization entanglement is largely employed as the prototypical example of a two-qubit entangled state. Several approaches for generating polarization entangled states have been developed, extending the capabilities of early SPDC sources that primarily produced entanglement in transverse position/momentum, or time/frequency. While the earliest methods  based on bulk type-II crystals were limited in efficiency due to the small collection probability~\cite{rubin1994theory, shih1994two,kwiat1995new}, modern approaches routinely generate polarization-entangled pairs with high brightness, using sources such as paired BiBO crystals~\cite{kwiat1999ultrabright}, periodically or aperiodically poled crystals~\cite{Steinlechner:12}, eventually inserted into interferometric setups~\cite{PhysRevA.73.012316,Fedrizzi:07,weston2016efficient}.

Beyond conventional polarization entanglement, newer schemes exploit spatially varying polarization entanglement (sometimes referred to as hyper-entanglement between spatial and polarization degrees of freedom), enabling the encoding of high-dimensional information with a limited number of quantum particles \cite{graffitti2020hyperentanglement,sit2021quantum, nemirovsky2024increasing}. For instance, polarization entanglement joined with spatial correlations has been used for holographic phase imaging~\cite{defienne2021polarization}, and supersensitive phase imaging across large fields of view \cite{camphausen2021quantum}. These works demonstrate that entanglement across multiple degrees of freedom can be harnessed to extract phase and structural information with sensitivity surpassing that of classical techniques.

Alongside imaging applications, several efforts have focused on quantifying and characterizing these complex, entangled states. Weak-measurement-based schemes have been proposed and demonstrated to extract Bell parameters from individual photon pairs without destroying the entanglement~\cite{genovese2025consequences,virzi2024entanglement}, while nonlocal weak measurements have been applied in microscopy to improve image contrast by filtering out environmental noise~\cite{liu2024metasurface}. Although powerful, these methods often rely on specialized interferometric arrangements and sequential projective measurements, making them less practical when fast, high-dimensional measurements are required.

In parallel, metasurfaces have emerged as a versatile platform for manipulating and characterizing quantum states of light. Dielectric metasurfaces have been used to generate and engineer entanglement~\cite{georgi2019metasurface}, induce quantum interference effects in the spin-orbit space of light~\cite{d2019tunable}, and to expand the capabilities of quantum state engineering by relaxing phase-matching constraints in nonlinear metasurfaces~\cite{santiago2022resonant,jia2025polarization}. Metasurfaces have also been designed for quantum measurements, including simultaneous projection of multiphoton states for robust tomography~\cite{wang2018quantum}, single-shot characterization of indistinguishability across multiple degrees of freedom~\cite{zhang2024single}, and complete measurement of all four polarization Bell states using binary-pixel metasurfaces~\cite{gao2023metasurface}. These works have emphasised the transformative potential of metasurfaces in nonclassical light generation, detection, and control \cite{solntsev2021metasurfaces}.
\begin{figure*}
    \includegraphics[width=0.9\textwidth]{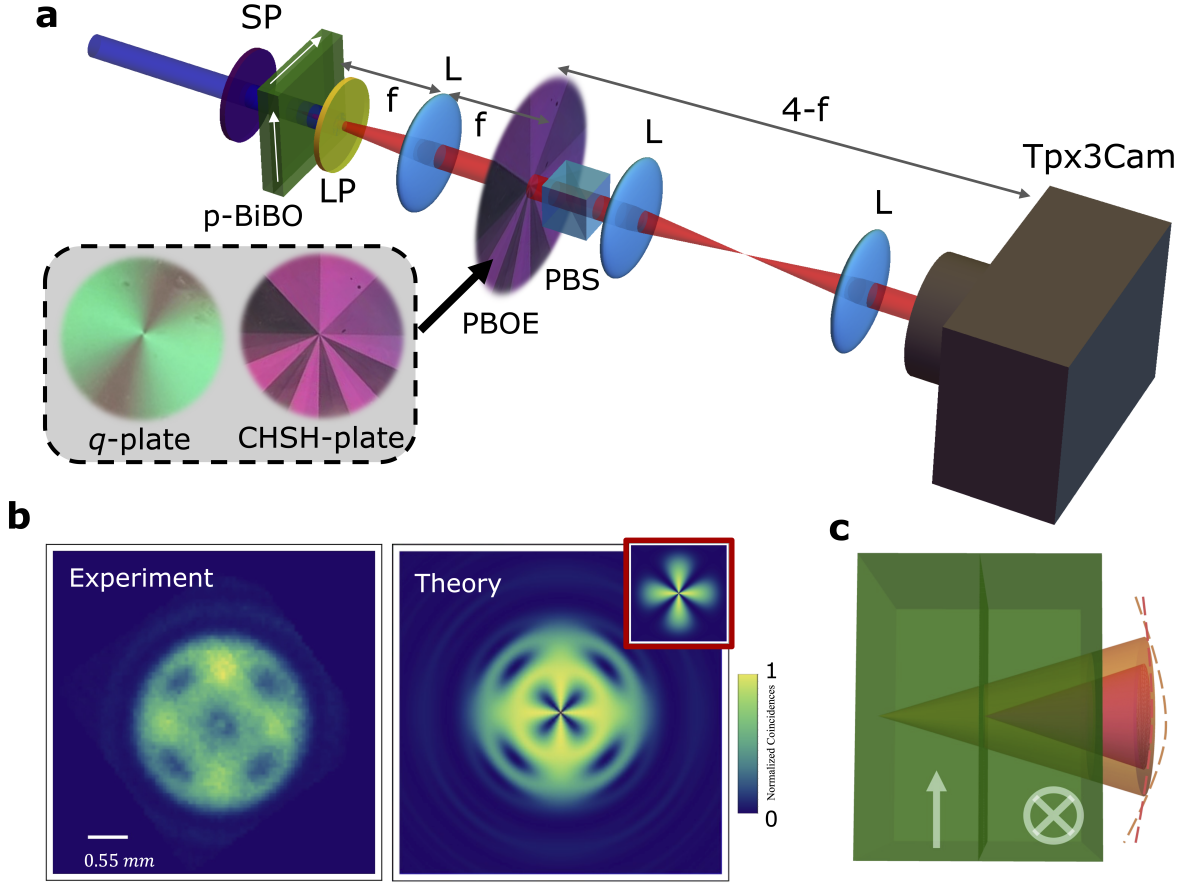}
\caption{ \textbf{Single-shot readout of structured entanglement with spin-orbit metasurfaces.} \textbf{(a)} Experimental implementation. A 355-nm pump laser illuminates a pair of BiBO crystals, generating polarization-entangled photon pairs by spontaneous parametric down-conversion (SPDC). The down-converted photons are sent through either a $q$-plate or a CHSH plate and are then imaged onto a time-stamping camera (Tpx3Cam), which records spatially resolved coincidence events. SP, short-pass filter; p-BiBO, paired BiBO crystals, LP, long-pass filter; L, lens; PBOE, Pancharatnam--Berry optical element. Photographs of the fabricated PBOEs between crossed polarizers are shown in the inset.  \textbf{(b)} Coincidence images after transmission through the $q$-plate, shown for the experiment (left) and the theoretical model (right). The four-fold azimuthal modulation arises from the spin-orbit transformation induced by the $q$-plate and carries information about both its topological charge and the relative phase of the two-photon amplitudes. The $q$-plate therefore maps the internal structure of the polarization-entangled state onto a directly measurable spatial interference pattern. \textbf{(c)} Origin of the spatial phase structure in the paired-crystal source. Photon pairs generated in the first crystal propagate through the second crystal before leaving the source, whereas pairs generated in the second crystal experience a shorter optical path. This imbalance introduces an additional quadratic phase in one contribution to the biphoton state, illustrated schematically by the dashed wavefronts, which is subsequently imprinted onto the entanglement distribution}
    \label{fig:fig1}
\end{figure*}
Despite these advances, a major challenge remains: efficient and rapid characterization of spatially varying polarization entanglement. States where different Bell states are associated with different positions \cite{fickler2014quantum,d2016entangled}  have been demonstrated in several works with the use of liquid crystal metasurfaces \cite{gao2024full} and quantum interference \cite{schiano2024engineering, gao2025generation}. Moreover, some sources, such as paired nonlinear crystals, naturally exhibit a spatial dependence of the polarization entanglement which can reduce the source quality if not compensated for \cite{defienne2021polarization,li2025rapid}. Most existing techniques rely on sequential measurements or full quantum-state reconstruction, both of which are time-intensive and scale poorly with system dimensionality. In many scenarios, however, it is sufficient to quantify entanglement directly, for example, through Bell inequality violations, without reconstructing the entire state. A method that enables such Bell measurements in a single shot would thus be highly valuable, particularly for applications in quantum imaging and sensing where speed and efficiency are critical. 


Previous works have demonstrated single-shot access to Bell-type correlations using spatially encoded qubits \cite{moreau2019imaging, mazelanik2021real}. Here, we address a fundamentally different and more general regime, in which the entanglement itself varies across the transverse plane and must be characterized locally. In this context, spatial correlations are not merely a resource for parallel readout, but provide the key mechanism for accessing position-dependent measurement contexts.

Specifically, we show that coupling polarization with the transverse wavevector, induced by Phancharatnam-Berry optical elements (PBOE), can transfer information encoded in polarization entanglement into the spatial degree of freedom. This results in interference patterns when spatially resolved coincidence measurements are performed. We demonstrate a liquid-crystal PBOE that performs spatially resolved Bell-inequality measurements in a single exposure. By exploiting the cylindrical symmetry of entangled states generated via SPDC, together with the transverse anti-correlations, our device applies different polarization projections across distinct azimuthal slices. When combined with an array of single-photon detectors, this allows us to map Bell parameters across the entire field in just one measurement, reducing the number of required exposures from 16 to 1.  This single-shot technique provides not only a practical tool for entanglement characterization but also a means to more efficiently explore the spatial structure and symmetries of entangled photon sources.
\section{Results}

Let us consider the polarization-entangled state generated by a pair of orthogonally oriented second-order nonlinear crystals \cite{kwiat1999ultrabright}. In the frequency-degenerate case, the two-photon state can be written as,
\begin{align}\label{eq:initialstate}
    \ket{\psi} &= \int d^2k_i\,d^2k_s\left(\Phi_1\,\hat{a}^{\dagger}_{H,\textbf{k}_i}\hat{a}^\dagger_{H,\textbf{k}_s}+\Phi_2\,\hat{a}^\dagger_{V,\textbf{k}_i}\hat{a}^\dagger_{V,\textbf{k}_s}\right)\ket{\text{vac}}
\end{align}
where $\hat{a}^{\dagger}{i}$ and $\hat{a}_{i}$ are the creation and annihilation operators of mode $i$, $H$ and $V$ are the horizontal and vertical polarization states, $\mathbf{k}_i$ and $\mathbf{k}_s$ denote the transverse momenta of the idler and signal photons, respectively, and $\Phi_1:=\Phi_1(\textbf{k}_i,\textbf{k}_s)$ and $\Phi_2:=\Phi_2(\textbf{k}_i,\textbf{k}_s)$ represent the biphoton wavefunctions associated with photon pairs produced in the first and second crystal, respectively. Photon pairs generated in the first crystal are horizontally polarized, while those produced in the second crystal are vertically polarized. Since pairs created in the first crystal must propagate through the second crystal before leaving the source, they acquire an additional quadratic phase relative to the pairs generated in the second crystal \cite{hegazy_tunable_2015, hegazy_relative-phase_2017}. 
The far-field entanglement distribution is thus well described by 
\begin{eqnarray}
\ket{\psi(\mathbf{k})}=&\Phi(k)e^{\frac{i}{2}\alpha(k)}\biggr(\cos{\left(\frac{\alpha(k)}{2}\right)}\ket{\vartheta^+}\cr\cr
&\,\,\,+i\sin{\left(\frac{\alpha(k)}{2}\right)}\ket{\vartheta^-}\biggr)\ket{\mathbf{k},-\mathbf{k}},\nonumber
\end{eqnarray}
where $\Phi(k)=|\Phi_1|=|\Phi_2|$, $\ket{\vartheta^{\pm}}=(\ket{H,H}\pm\ket{V,V})/\sqrt{2}$, $k:=\abs{\mathbf{k}}$, and $\alpha(k)=\arg(\Phi_1)-\arg(\Phi_2)\propto k^2$.
This phase structure has been previously reconstructed using full spatial quantum-state tomography~\cite{defienne2021polarization,li2025rapid}, which requires a series of spatially resolved measurements. As we show here, however, the presence of this phase can be revealed more directly using spatially varying birefringent elements.

\subsection{Spatially Varying Polarization Entanglement Probed by a q-plate}
By introducing a spin-orbit metasurface, such as a $q$-plate, into the optical path, the phase difference between $\Phi_1$ and $\Phi_2$ gives rise to a measurable modulation in the coincidence distribution of the photon pairs.
The spin-orbit metasurface used here is a $q$-plate, a liquid-crystal device in which the orientation of the optical axis varies with the azimuthal coordinate $\phi$ according to $\theta = q\phi$, with $2q\in\mathbb{Z}$ being an integer number~\cite{rubano2019q}. As a result, the plate behaves as a spatially varying wave plate: each point in the transverse plane applies a polarization transformation determined by the local liquid-crystal alignment. The strength and symmetry of this transformation are set by the topological charge $q$. When placed in the far field of the photon-pair source, the $q$-plate effectively converts polarization information into spatial structure. After projecting both photons onto horizontal polarization, the device can therefore be interpreted as implementing spatially dependent polarization projections of the entangled state. The experimental implementation is shown in Fig.~\ref{fig:fig1}-\textbf{a}. The entangled photons were generated in a paired BiBO crystal pumped by a 355\,nm XCyte laser, producing degenerate photon pairs at 710\,nm. The resulting spatially resolved coincidence patterns were recorded with a time-stamping camera (Tpx3Cam), which allows direct reconstruction of the transverse correlation distribution.

The transformation induced by a $q$-plate can be conveniently expressed in the circular polarization basis. Defining the creation-operator vector $\hat{\mathbf{a}}^{\dagger} = \begin{pmatrix} \hat{a}_L & \hat{a}_R \end{pmatrix}^{\dagger}$, the output operators are given by $\hat{\mathbf{b}}^{\dagger}=\hat{U}(\delta,q,\phi)\,\hat{\mathbf{a}}^{\dagger},$ where the operator describing the $q$-plate is
\begin{eqnarray}\label{eq:$q$-plate}
\hat{U}(\delta,q,\phi)
=
\cos\!\left(\frac{\delta}{2}\right)\mathbb{I}
+i\,\sin\!\left(\frac{\delta}{2}\right)
\begin{pmatrix}
0 & e^{i2q\phi} \\
e^{-i2q\phi} & 0
\end{pmatrix}.
\end{eqnarray}
$\mathbb{I}$ denotes the $2\times2$ identity matrix, and $\delta$ is the optical retardation of the $q$-plate, which in our experiment can be controlled by an external AC electric field. Here, we consider the effect of a tuned $q$-plate, $\delta=\pi$, placed in the far field of the crystal on the generated entangled state of Eq.~\eqref{eq:initialstate}. Assuming a plane-wave pump laser, that implies sharp spatial anti-correlations in the far field, the two photon wavefunctions can be simplified to $\Phi_{1,2}(\textbf{k}_i,\textbf{k}_s)\rightarrow\Phi_{1,2}(\textbf{k}_i,-\textbf{k}_i)=:\Phi_{1,2}(\textbf{k})$. Applying the transformation of Eq.~\eqref{eq:$q$-plate} to the state in Eq.~\eqref{eq:initialstate}, and projecting both photons onto horizontal polarization, the resulting biphoton wavefunction can be written as (see the Supplementary Information S1 for details)
\begin{equation}\label{eq:totalbiphoton}
	\widetilde\psi(\textbf{k})=\left(\Phi_1(k)+\Phi_2(k)\right)\text{cos}(4q\phi)+\left(\Phi_1(k)-\Phi_2(k)\right)
\end{equation}
where $k=\lvert\mathbf{k}\rvert$. 
The expression in Eq.~\eqref{eq:totalbiphoton} follows from the spin-orbit transformation imposed by the $q$-plate. 
In the circular polarization basis, the device couples spin and orbital angular momentum (OAM) by introducing an azimuthally varying phase factor $\exp(\pm i2q\phi)$ between the two helicity components. When the output photons are projected onto horizontal polarization, the resulting amplitude contains contributions from the initial polarization state of each photon before the metasurface, expressed in the circular basis, together with the azimuthal phase acquired by each circular component after the metasurface. Consequently, all four initial polarization combinations ($RR,\,RL,\,LR$, and $LL$), contribute to the azimuthal modulation in Eq.~\eqref{eq:totalbiphoton}. Physically, this modulation arises because the metasurface converts polarization correlations into spatial interference. As a result, information about the relative phase between the two biphoton amplitudes $\Phi_1$ and $\Phi_2$ becomes directly visible in the transverse coincidence distribution.

When the $q$-plate is aligned with the centre of the SPDC cone, the measured coincidence distribution, proportional to $\lvert\widetilde{\psi}(\mathbf{k})\rvert^2$, exhibits a characteristic $8q$-fold modulation (see Fig.~\ref{fig:fig1}-\textbf{b} for $q=1/2$). This pattern reflects the spatial structure of the entangled state and follows directly from Eq.~\eqref{eq:totalbiphoton}, where the $\cos(4q\phi)^2$ term produces an azimuthal modulation. In the ideal case $\Phi_1=\Phi_2$, this four-fold pattern would appear with maximal visibility and remain independent of the radial coordinate. The radial variation observed experimentally, therefore, originates from the second term in Eq.~\eqref{eq:totalbiphoton}, which reflects the phase difference between the two contributions to the biphoton state. To describe this behaviour, we assume that the two crystals are otherwise identical, such that the radial modulation arises solely from the small difference in propagation distance experienced by photon pairs generated in the first crystal. As illustrated in Fig.~\ref{fig:fig1}-\textbf{c}, this imbalance introduces a curvature in the phase-matching wavefront associated with the first contribution~\cite{li2025rapid,dehghan2024biphoton}. Following the standard phase-matching description of SPDC \cite{walborn2010spatial}, we model the amplitudes as
\begin{equation}
    \abs{\Phi_1}=\abs{\Phi_2}=\mathrm{sinc}(8\pi k^2/L\lambda_p-\xi)
\end{equation}
and introduce a relative phase 
\begin{equation}
    \arg(\Phi_1)-\arg(\Phi_2)=-k^2/R^2.
\end{equation}

Here $L$ is the crystal length ($0.5\,\mathrm{mm}$), $\lambda_p$ is the pump wavelength, $\xi$ denotes the longitudinal walk-off parameter, and $R$ represents an effective curvature radius used as a fitting parameter. This curvature is imprinted directly in the entanglement distribution, as exemplified by the experimental measurement of the CHSH parameter, described in the next section.

As shown in the right plot of Fig.~\ref{fig:fig1}-\textbf{b}, the model reproduces the main features of the measured coincidence patterns after the $q$-plate action. In particular, the radial modulation arises from quantum interference between the two biphoton contributions, which acquire slightly different wavefront curvatures due to their distinct propagation paths -- in the absence of the radial phase curvature the coincidence distribution would be as shown in the inset. This behaviour is consistent with observations reported in previous studies~\cite{defienne2021polarization,li2025rapid}. More generally, these results illustrate how a $q$-plate can serve as a simple and efficient tool for probing the spatial distribution of polarization entanglement.

\subsection{Single-Shot Violation of the CHSH Inequality Using Liquid Crystal Metasurfaces}
While the scheme above reveals the spatial phase structure of the biphoton state, it does not by itself provide a direct test of entanglement, such as a Bell inequality violation. To verify that the observed radial modulation indeed originates from spatially varying polarization entanglement, we performed a radially-resolved CHSH test.  

To this end, we devised a measurement strategy based on the idea that the transverse coordinate of the detection event does not merely label the spatial position of the photon pair but specifies the measurement context itself.  Signal and idler photons produced by SPDC exhibit strong transverse momentum anti-correlations, such that detection of one photon at azimuthal angle $\phi$ implies that its partner photon is detected near the diametrically opposite position $\phi+\pi$. We developed the CHSH metasurface, which exploits this property by implementing different polarization projections in distinct azimuthal sectors. Consequently, each coincidence event occurring in a pair of opposite sectors corresponds to a well-defined joint polarization measurement $(\alpha_i,\beta_j)$. In this configuration, the metasurface assigns different measurement settings to different spatial regions of the beam. The azimuthal coordinate, therefore, acts as a classical register selecting the measurement context, while the photon pair samples one of these contexts according to its emission direction. Operationally, the CHSH metasurface implements a spatially multiplexed measurement described by the operator
\begin{equation}\label{eq:measurement}
	\hat{M} = \int d\phi \, \hat{\Pi}_{A}(\phi)\otimes \hat{\Pi}_{B}(\phi+\pi)\otimes |\phi,\phi+\pi\rangle\langle \phi,\phi+\pi|,
\end{equation}
where $\hat{\Pi}_{A}(\phi)$ and $\hat{\Pi}_{B}(\phi+\pi)$ denote polarization projectors defined by the local optic-axis orientation of the metasurface. The spatial degree of freedom, therefore, acts as a label of the measurement context, while the coincidence detection selects the corresponding joint projection. The approximate cylindrical symmetry of the source implies that the polarization correlations vary predominantly along the radial coordinate.

\begin{figure*}
    \includegraphics[width=2\columnwidth]{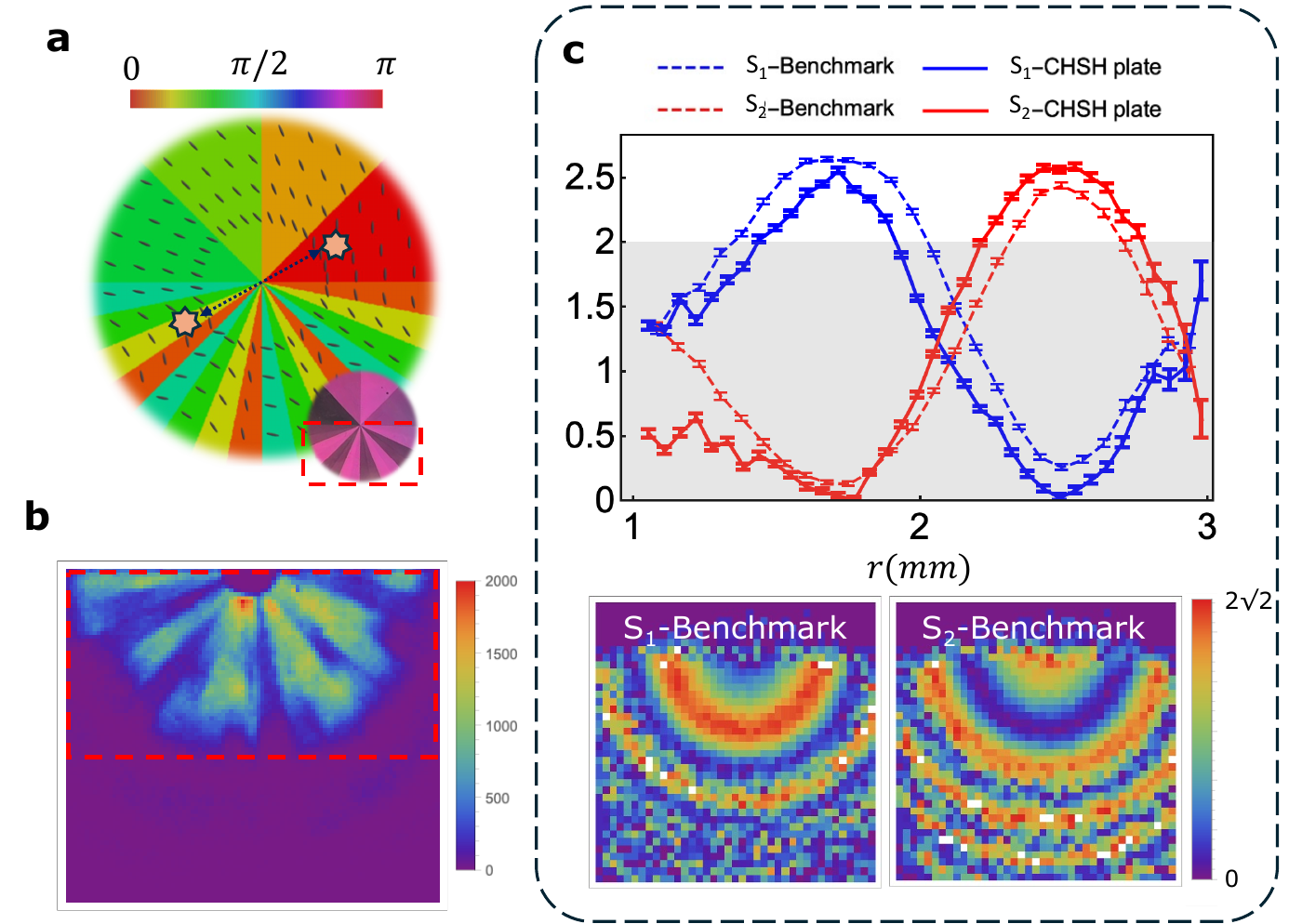}
\caption{\textbf{Single-shot CHSH measurement with the metasurface.} \textbf{(a)} Spatial pattern of the optic-axis orientation in the CHSH plate, which encodes the different polarization projections across the transverse plane. \textbf{(b)} Coincidence image of the SPDC photons after transmission through the CHSH plate. Different azimuthal sectors correspond to projections onto distinct linear polarization states, leading to characteristic variations in the photon distribution around the ring. \textbf{(c)} Comparison between the CHSH parameters extracted from the azimuthal sectors in \textbf{b} using a single measurement and those obtained with the benchmark setup based on 16 sequential polarization measurements, as seen in the insets. The benchmark values are averaged over all azimuthal angles. To obtain similar error-bars for the benchmark data and single shot data the exposure time corresponding to panel (b) was chosen to be approximately 10-times longer than for the single projections in the benchmark measurements.}
    \label{fig:fig3}
\end{figure*}
From the sector-resolved coincidence statistics, we evaluate the correlators \cite{aspect1981experimental, aspect1982experimental,PhysRevLett.49.91},
\begin{equation}\label{eq:correlator}
	E_{\alpha_i,\beta_j}(r)=\frac{N_{++}-N_{+-}-N_{-+}+N_{--}}{N_{++}+N_{+-}+N_{-+}+N_{--}},
\end{equation}
where each $N$ is the number of counts accumulated in each projective measurement. $``+"$ corresponds to projections along linearly polarized states oriented along $\alpha$ or $\beta$, and $``-"$ corresponds to projections along states oriented along the perpendicular direction to $\alpha$ or $\beta$ ($\bar{\alpha}=\alpha+\frac{\pi}{2}$ or $\bar{\beta}=\beta+\frac{\pi}{2}$) --- a similar notation holds for $\alpha'$ and $\beta'$. From these correlations, the local CHSH parameters can be obtained,
\begin{equation}\label{eq:chsh-parameters}
\begin{split}
    &S_1(r) = -E_{\alpha,\beta}(r)+E_{\alpha,\beta'}(r)+E_{\alpha',\beta}(r)+E_{\alpha',\beta'}(r),\\
    &S_2(r) =E_{\alpha,\beta}(r)-E_{\alpha,\beta}(r)+E_{\alpha',\beta}(r)+E_{\alpha',\beta'}(r), 
\end{split}
\end{equation}
where $\alpha= 0$, $\beta = \frac{\pi}{8}$, $\alpha' = \frac{\pi}{4}$ and $\beta'=\frac{3\pi}{8}$, correspond to projection angles.

In any non-contextual, measurement-independent, hidden-variable model, the outcomes assigned to local polarization observables are assumed to be independent of the jointly implemented measurement context. Such models therefore impose the bound $|S_{i}(r)|\le2$ for every radial coordinate $r$. Quantum mechanics predicts instead
\begin{equation}\label{eq:chsh-value}
	S_i(r)=2\sqrt{2}\,V(r),
\end{equation}
where $V(r)$ is the local interference visibility of the entangled state, leading to violations whenever $V(r)>1/\sqrt{2}$. In this way, the metasurface converts the usual requirement of actively switching analyzer settings in Bell experiments into a passive spatial encoding of measurement contexts. 
Each photon pair samples one context according to its emission direction, while the full set of contexts is realized simultaneously across the transverse plane. The observed violations of the CHSH bound therefore demonstrate that the sector-resolved correlations cannot be reproduced by any local hidden-variable model satisfying measurement independence, and are consistent with entangled two-photon correlations implemented through this spatially multiplexed measurement.


\begin{figure*}
\includegraphics[width=0.9\textwidth]{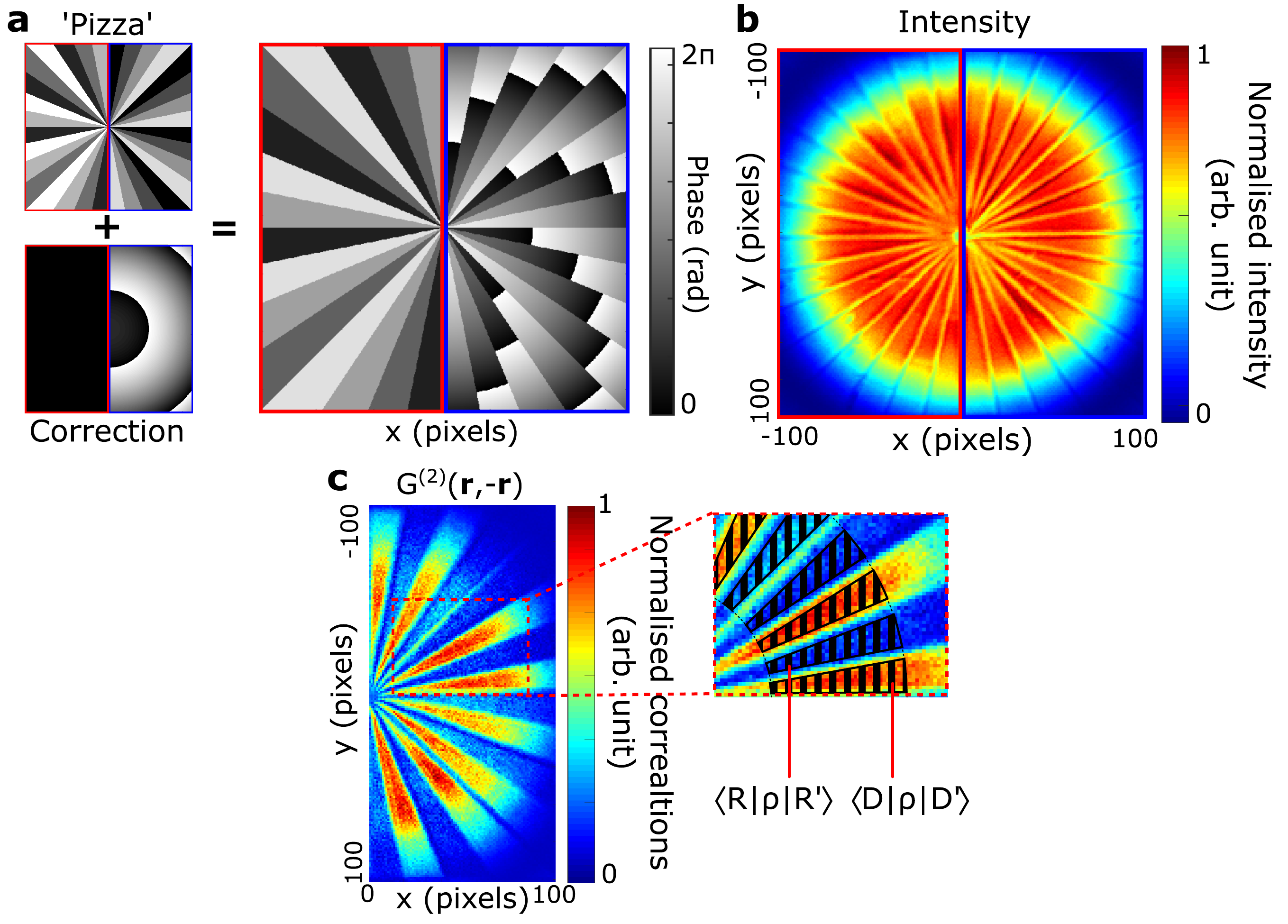}
\caption{\textbf{Programmable CHSH measurement using spatial light modulators.} \textbf{(a)} Phase mask applied to the SLM for the CHSH measurement. The pattern combines the angular ``pizza'' phase mask, which implements the 16 joint polarization projections, with the phase correction $\Psi(x, y) = 4.69k_x^2 + 5.04k_y^2 + 0.02$ that compensates for the spatial phase distortion of the biphoton state. \textbf{(b)} Intensity distribution of the SPDC field recorded with the SLM pattern applied. Diffraction features appear at the boundaries between adjacent angular sectors. \textbf{(c)} Spatially resolved correlation image $G^{(2)}(\boldsymbol{r},-\boldsymbol{r})$ obtained with the EMCCD camera. Each pair of centrosymmetric sectors corresponds to one of the 16 joint polarization measurements required to evaluate the CHSH parameter. To avoid diffraction artefacts at the sector boundaries, the CHSH value is computed from data in the interior regions of each slice.} 
\label{fig:fig4}
\end{figure*}


To evaluate the CHSH parameters, each photon must be projected onto four distinct linear polarization states. This requirement can be implemented by partitioning one half of the cone into four azimuthal sectors of equal angular width $\pi/4$, each corresponding to a different polarization projection (see Fig.~\ref{fig:fig3}-\textbf{a}). The diametrically opposite regions are then subdivided into four additional sectors associated with the complementary projections of the partner photon. In this way, each coincidence event corresponds to one of the 16 joint polarization measurements required for the CHSH test, effectively dividing the cone into $16$ azimuthal sectors of angular width $\pi/16$ (Fig.~\ref{fig:fig3}-\textbf{a,b}). This projection scheme is implemented using a spin-orbit metasurface, similar in spirit to a $q$-plate but with a specifically engineered azimuthal dependence of the optic-axis orientation $\theta(\phi)$. By tailoring the liquid-crystal alignment at each azimuthal position, the metasurface locally defines the required polarization projection. When placed in the far field of the source, this device enables all 16 joint measurements to be recorded simultaneously. We refer to this element as a \textit{CHSH plate}. 

Figure~\ref{fig:fig3}-\textbf{a} shows the corresponding optic-axis pattern $\theta(x,y)$, together with an image of the fabricated device viewed between crossed polarizers. 
The CHSH plate implements a spatially multiplexed Bell measurement in which incompatible measurement contexts coexist across the beam profile. The coincidence image therefore, directly contains the outcomes of all CHSH measurements. After recording coincidence events and post-selecting photon pairs exhibiting sharp transverse momentum anti-correlations \cite{zia2023interferometric,dehghan2024biphoton}, we obtain a correlation image that simultaneously contains the results of all 16 joint polarization projections (Fig.~\ref{fig:fig3}-\textbf{b}). Each azimuthal slice of the image corresponds to one of the required CHSH measurement settings. By averaging the coincidence counts within each slice, for a fixed radius $r$, we extract the corresponding measurement outcomes as a function of the radial coordinate. The Bell parameters $S_1$ and $S_2$ are then calculated using the same procedure employed in the conventional sequential measurements, as given by Eq.~\eqref{eq:chsh-parameters}. 

Figure~\ref{fig:fig3}-\textbf{c} compares the CHSH parameters obtained from the conventional benchmark measurements (dashed curves), which require 16 sequential polarization projections (see Supplementary Materials) with those obtained from the single-shot measurement performed using the CHSH plate (solid curves). For the benchmark data shown in the insets of Fig.\ref{fig:fig3}-\textbf{c}, the plotted values correspond to azimuthally averaged results. Overall, the two approaches show good qualitative agreement. Residual discrepancies between the two methods can arise from several experimental imperfections. In particular, small misalignments between the CHSH plate and the centre of the SPDC cone can lead to imperfect photon splitting and reduced coincidence counts. Additional deviations may result from spatial variations in the retardation of the liquid-crystal device, which is fabricated in-house and can exhibit small thickness inhomogeneities across the aperture. Although the device was calibrated using projective measurements to determine the voltage corresponding to $\pi$ retardation, minor non-uniformities remain unavoidable. We also observe larger deviations at small radii, where cross-talk between neighbouring azimuthal slices becomes more significant. These effects depend on both the alignment of the metasurface and the finite width of the far-field spatial correlations. 

We further examined the robustness of the measurement by slightly shifting the CHSH plate along the optical axis by a few millimetres. The Bell inequality violation remained clearly observable, although the region of violation became smaller at small radii (see Supplementary Materials). This reduction is consistent with the broadening of the spatial anti-correlations when the metasurface is displaced from the optimal imaging plane. These observations highlight the  tolerance for moderate longitudinal misalignment. The CHSH plate provides a convenient transmissive object to characterize entangled photon sources in a compact and simple setup. In the following section we show that more general actions can be performed with reconfigurable reflective spatial light modulators (SLMs).

\subsection{Programmable Single-Shot CHSH Measurement with Phase Compensation Using SLMs}


To demonstrate the flexibility of the spatially multiplexed measurement approach, we implement a programmable version of the scheme using a spatial light modulator (SLM). In this configuration, the transverse momentum of the SPDC photons is mapped onto the SLM plane, where spatially varying phase shifts are applied to the horizontal polarization component of each spatial mode. A polarizer placed after the SLM converts these phase shifts into polarization projections, so that each device pixel implements a specific projective measurement on the polarization state associated with the corresponding transverse momentum (see Supplementary Materials for experimental details). By encoding an angular “pizza”-like phase pattern, the SLM realizes the set of 16 joint measurements required for evaluating the CHSH inequality within a single acquisition. In contrast to the static metasurface, this approach allows dynamic reconfiguration of the measurement basis. 

A key advantage of the programmable implementation is the ability to compensate for residual spatial phase distortions in the biphoton state associated with the crystal geometry. We characterize this distortion using a quantum holography technique and incorporate the corresponding correction directly into the SLM phase mask. This compensation ensures uniform measurement bases across the transverse plane and improves the accuracy of the extracted correlations. Using the corrected phase pattern, we obtain a spatially resolved correlation image from which the CHSH parameter is directly evaluated, yielding $S=2.56\pm0.16$, in clear violation of the classical bound. Representative phase masks and the corresponding measurement results are shown in Fig.~\ref{fig:fig4} (further details shown in the Supplementary Materials).\newline

\section{Conclusions and discussion.}
In conclusion, we have demonstrated that Bell measurements can be spatially multiplexed by encoding measurement contexts directly in the transverse structure of an optical element. Using a $q$-plate, we demonstrated that the relative wavefront curvature of biphoton states generated by a pair of nonlinear crystals can be retrieved from the spatial structure of the coincidence signal. Building on this principle, we developed a liquid-crystal metasurface, the CHSH plate, that enables spatially multiplexed Bell measurements, allowing the 16 joint polarization projections required in a conventional CHSH test to be implemented simultaneously across the beam's transverse profile. In this architecture, the azimuthal coordinate of the SPDC cone effectively encodes the measurement context, while photon pairs randomly sample these contexts according to their emission direction. As a result, the Bell inequality can be evaluated in a single acquisition rather than by sequentially switching analyzer settings. We further demonstrated that the same measurement concept can be implemented using a programmable spatial light modulator, providing a dynamic realization in which the measurement contexts are defined by spatially varying phase shifts. In this sense, the metasurface used in our experiments can be viewed as a static implementation of a more general programmable measurement architecture. 

Beyond its experimental realization, our work establishes a distinct measurement paradigm in which the spatial coordinate does not merely label different photon pairs, but defines the measurement context itself. Different measurement settings coexist across the transverse plane, and each photon pair samples one of them according to its emission direction. The observed Bell violations are therefore interpreted, in the standard CHSH framework, as evidence of two-photon polarization entanglement whose measurement settings are spatially multiplexed across the transverse plane.

By eliminating the need for sequential measurements and reducing experimental overhead, this approach provides a powerful tool for probing the spatial structure and symmetries of entangled photon sources. In particular, it enables rapid characterization of spatially structured entanglement and may prove useful for studying photon pairs generated by metasurfaces or other engineered nonlinear platforms. More broadly, the ability to encode measurement contexts directly in the spatial geometry of an optical device opens new possibilities for parallel quantum measurements, entanglement-based quantum key distribution, Bell-inequality-based imaging, and quantum-enhanced parameter estimation.

\bibliography{bibliography}

\subsection*{Funding}
This work was supported by the Canada Research Chairs (CRC), the NRC Quantum Sensing Programme, Quantum Enhanced Sensing, Imaging (QuEnSI) Alliance Consortia Quantum grant. 
\subsection*{Author contributions} E.K., A.D., and N.D. conceived the idea, devised the experiment and prepared the setup. E.K. fabricated the $q$-plate. N.D., with contributions from A.D., devised and fabricated the CHSH-plate and collected the data. H.D. and D.F. conceived the spatial light modulator idea, conducted the experiment and analysed the data. Y.Z. prepared the software for the readout and analysis of the Tpx3Cam data. N.D. analysed the data and, with contributions from A.D., prepared the first version of the manuscript.  All authors contributed to the revision of the manuscript.

\subsection*{Competing interest}
The authors declare that they have no competing interests.
\subsection*{Data and materials availability}
All data are available in the main text or the supplementary materials.

\clearpage
\onecolumngrid
\renewcommand{\figurename}{\textbf{Figure}}
\setcounter{figure}{0} \renewcommand{\thefigure}{\textbf{S{\arabic{figure}}}}
\renewcommand{\theHfigure}{S\arabic{figure}}
\setcounter{table}{0} \renewcommand{\thetable}{S\arabic{table}}
\setcounter{section}{0} \renewcommand{\thesection}{S\arabic{section}}
\setcounter{equation}{0} \renewcommand{\theequation}{S\arabic{equation}}
\onecolumngrid
\begin{center}
{\Large Supplementary Material for: \\Single-shot Detection of Structured Entanglement with Spin-Orbit Metasurfaces}
\end{center}
\vspace{1 EM}

\section{Wavefunction Calculation} 

We start from the state defined in Eq.~\eqref{eq:initialstate} and derive the final form of the wavefunction Eq.~\eqref{eq:totalbiphoton}. To do so, let us begin by expanding $a^\dagger_{H}$ and $a^\dagger_V$ in terms of $a^\dagger_{L}$ and $a^\dagger_{R}$. For simplicity, we will omit the normalization factors. 
\begin{equation}
\begin{split}
    \psi &= \Phi_1a^\dagger_{H,\textbf{k}_i}a^\dagger_{H,\textbf{k}_s}+\Phi_2a^\dagger_{V,\textbf{k}_i}a^\dagger_{V,\textbf{k}_s}\\&=\Phi_1(a^\dagger_{L,\textbf{k}_i}+a^\dagger_{R,\textbf{k}_i})(a^\dagger_{L,\textbf{k}_s}+a^\dagger_{R,\textbf{k}_s})+\Phi_2(a^\dagger_{L,\textbf{k}_i}-a^\dagger_{R,\textbf{k}_i})(a^\dagger_{L,\textbf{k}_s}-a^\dagger_{R,\textbf{k}_s})\\&=(\Phi_1+\Phi_2)(a^\dagger_{L,\textbf{k}_i}a^\dagger_{L,\textbf{k}_s}+a^\dagger_{R,\textbf{k}_i}a^\dagger_{R,\textbf{k}_s})+(\Phi_1-\Phi_2)(a^\dagger_{L,\textbf{k}_i}a^\dagger_{R,\textbf{k}_s}+a^\dagger_{R,\textbf{k}_i}a^\dagger_{L,\textbf{k}_s}). 
\end{split}
\end{equation}
Now, using the action of the $q$-plate from Eq.~\eqref{eq:$q$-plate}, and defining $\psi_{+}=\Phi_1+\Phi_2$ and $\psi_{-}=\Phi_1-\Phi_2$, we can expand our wavefunction as follows,
\begin{equation}
    \begin{split}
        \psi'=&\psi_{+}[(\cos{\frac{\delta}{2}}a^\dagger_{L,\textbf{k}_i}+i \sin{\frac{\delta}{2}} e^{-i2q\phi_i} a^\dagger_{R,\textbf{k}_i})(\cos{\frac{\delta}{2}}a^\dagger_{L,\textbf{k}_s}+i \sin{\frac{\delta}{2}} e^{-i2q\phi_s} a^\dagger_{R,\textbf{k}_s})\cr&+(i\sin{\frac{\delta}{2}}e^{i2q\phi_i} a^\dagger_{L,\textbf{k}_i}+\cos{\frac{\delta}{2}}a^\dagger_{R,\textbf{k}_i})(i\sin{\frac{\delta}{2}}e^{i2q\phi_s} a^\dagger_{L,\textbf{k}_s}+\cos{\frac{\delta}{2}}a^\dagger_{R,\textbf{k}_s})]\\&+\psi_{-}[(\cos{\frac{\delta}{2}}a^\dagger_{L,\textbf{k}_i}+i \sin{\frac{\delta}{2}} e^{-i2q\phi_i} a^\dagger_{R,\textbf{k}_i})(i\sin{\frac{\delta}{2}}e^{i2q\phi_s} a^\dagger_{L,\textbf{k}_s}+\cos{\frac{\delta}{2}}a^\dagger_{R,\textbf{k}_s})\\&+(i\sin{\frac{\delta}{2}}e^{i2q\phi_i} a^\dagger_{L,\textbf{k}_i}+\cos{\frac{\delta}{2}}a^\dagger_{R,\textbf{k}_i})(\cos{\frac{\delta}{2}}a^\dagger_{L,\textbf{k}_s}+i \sin{\frac{\delta}{2}} e^{-i2q\phi_s} a^\dagger_{R,\textbf{k}_s})].
    \end{split}
\end{equation}
Finally, projecting both photons onto Horizontal polarization, we get,
\begin{equation}
    \begin{split}
        \psi'=&\psi_{+}[\cos^2{\frac{\delta}{2}}-\sin^2{\frac{\delta}{2}}\cos{2q(\phi_i+\phi_s)}+i\cos{\frac{\delta}{2}}\sin{\frac{\delta}{2}}(\cos{2q\phi_i}+\cos{2q\phi_s})]\\&+\psi_{-}[\cos^2{\frac{\delta}{2}}-\sin^2{\frac{\delta}{2}}\cos{2q(\phi_i-\phi_s)}+i\cos{\frac{\delta}{2}}\sin{\frac{\delta}{2}}(\cos{2q\phi_i}+\cos{2q\phi_s})].
    \end{split}
\end{equation}
Now, considering that in the farfield of the SPDC state, due to the anti-correlations, we can replace $\phi_i$ with $\phi_s+\pi$, therefore,
\begin{equation}
    \begin{split}
        \psi'=&\psi_{+}[\cos^2{\frac{\delta}{2}}-\sin^2{\frac{\delta}{2}}\cos{(4q\phi+2q\pi)}+i\cos{\frac{\delta}{2}}\sin{\frac{\delta}{2}}(\cos{2q(\phi+\pi)}+\cos{2q\phi})]\\&\psi_{-}[\cos^2{\frac{\delta}{2}}-\sin^2{\frac{\delta}{2}}(-1)^{2q})+i\cos{\frac{\delta}{2}}\sin{\frac{\delta}{2}}(\cos{2q(\phi+\pi)}+\cos{2q\phi})].
    \end{split}
\end{equation}
In our experiment, we set $\delta=\pi$, which will simplify our equation to
\begin{equation}
    \psi'=(\Phi_1+\Phi_2)\cos{(4q\phi)} + (\Phi_1-\Phi_2),
\end{equation}
that is the same equation as Eq.~\eqref{eq:totalbiphoton}.

\begin{figure*}
    \includegraphics[width=0.8\textwidth]{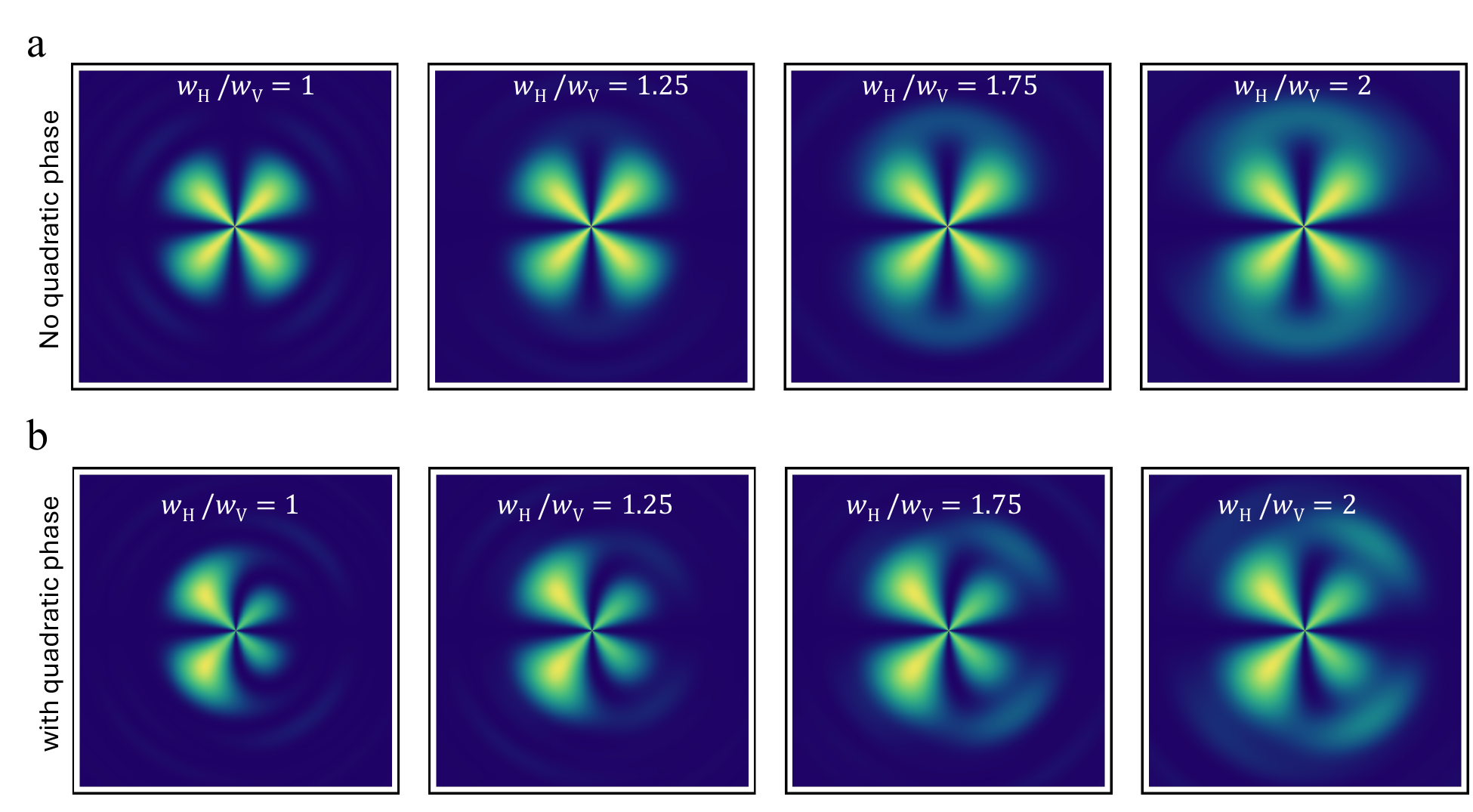}
    \caption{Simulations of coincidence patterns emerging when an input state of the form $\ket{H,V}+e^{i\alpha(|k|)}\ket{V,H}$ is input in a $q$-plate with $q=1/2$ and $\delta=\pi/2$.}
    \label{fig:Simulations}
\end{figure*}

The same calculation can be done for an initial state of the form,
\begin{equation}
    \ket{\psi} = \int d^2k_i\,d^2k_s\left(\Phi_1(\textbf{k}_i,\textbf{k}_s)a^\dagger_{H,\textbf{k}_i}a^\dagger_{V,\textbf{k}_s}+\Phi_2(\textbf{k}_i,\textbf{k}_s)a^\dagger_{V,\textbf{k}_i}a^\dagger_{H,\textbf{k}_s}\right)\ket{vac}.
    \label{eq:initialstateHVVH}
\end{equation}
which can be obtained in Type-II periodically poled crystals.
Going through the exact same procedure, the final result would be,
\begin{equation}
        \psi'=\psi_{+}\left(i\sin^2{\frac{\delta}{2}}\sin{(4q\phi+2q\pi)}+\cos{\frac{\delta}{2}}\sin{\frac{\delta}{2}}(\sin{2q(\phi+\pi)}+\sin{2q\phi})\right)+\psi_{-}\left(\cos{\frac{\delta}{2}}\sin{\frac{\delta}{2}}(\sin{2q(\phi+\pi)}-\sin{2q\phi})\right).
\end{equation}
In this case, while $\delta=\pi$ does not yield a quantum interference effect, choosing $\delta=\pi/2$ will allow to observe the interference pattern given by the square modulus of the function
\begin{equation}
    \psi'=-\frac{i}{2}(\Phi_1+\Phi_2)\sin{(4q\phi)} - (\Phi_1-\Phi_2) \sin{(2q\phi)}.
\end{equation}
The resulting interference patterns are shown in Fig.~\ref{fig:Simulations} for two different situations. In Fig.~\ref{fig:Simulations}-a we consider the case that $\Phi_1$ and $\Phi_2$ have the same functional form $\text{sinc}(k^2/w^2-\xi)$ with different characteristic width parameters $w_{H,V}$. This situation can effectively model the output of Type-II ppKTP crystals with imperfect temperature tuning. We also consider in Fig.~\ref{fig:Simulations}-b the case where an additional quadratic phase is applied on $\Phi_2$ (as it could be observed in Sagnac based sources with misplaced crystal position). 

\section{Experimental setup for benchmark measurement}

\begin{figure*}
\includegraphics[width=0.8\textwidth]{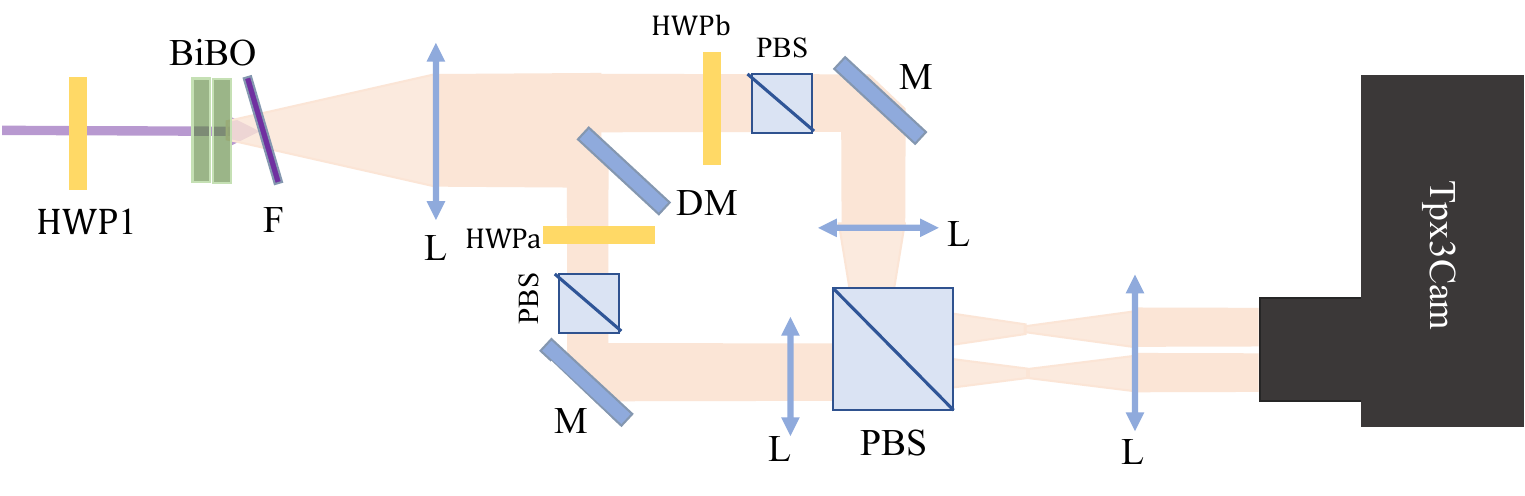}
\caption{Experimental setup for benchmark CHSH measurements. BiBO is a paired BiBO crystal. HWP1 is used to prepare the pump in a diagonal polarization state. F is a high-pass filter. L: lens, M: mirror, PBS: polarizing beam splitter, HWPa and HWPb: waveplates for state projections, DM: D-shaped mirror.}
\label{fig:S2}
\end{figure*}

\begin{figure*}
    \includegraphics[width=0.7\columnwidth]{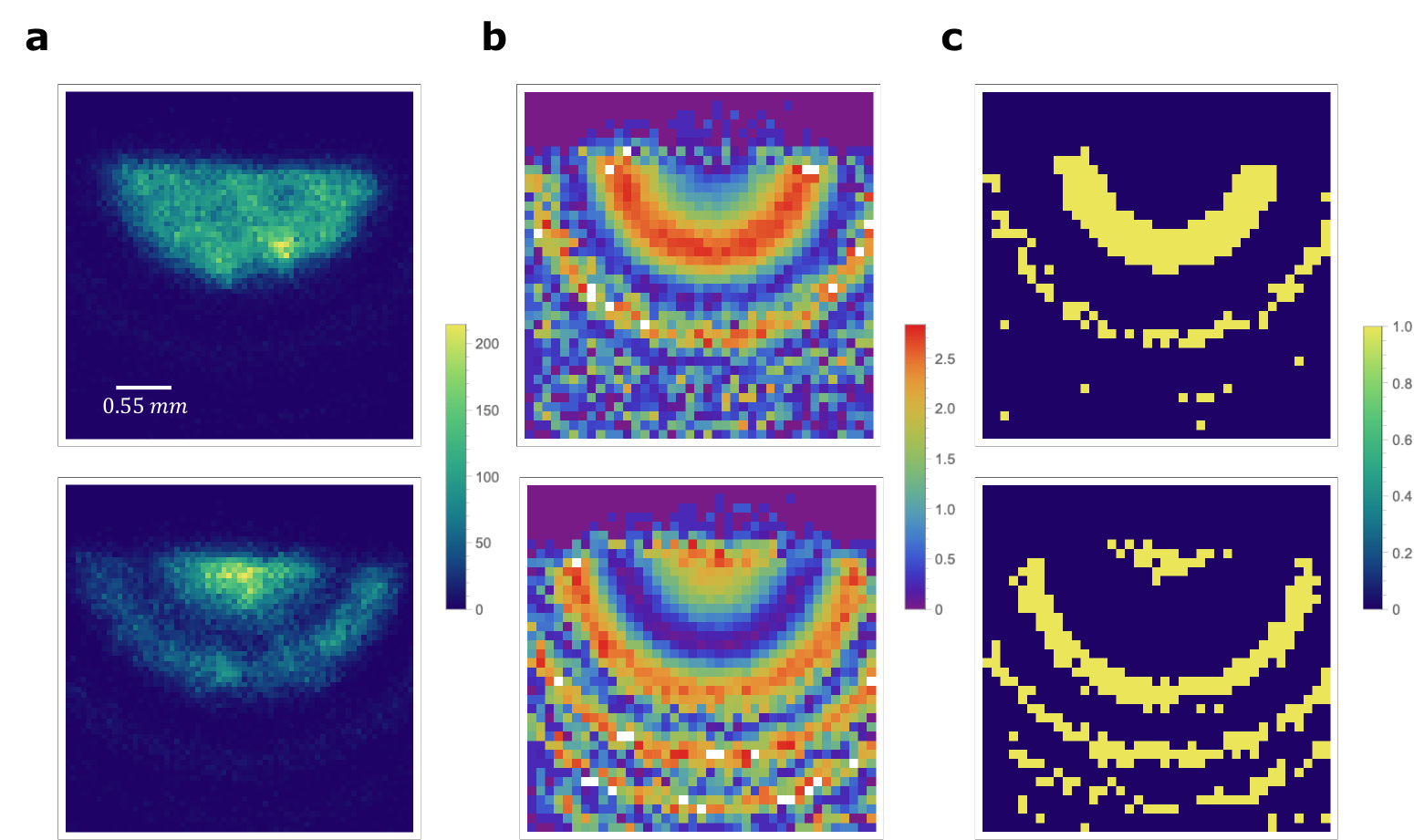} 
    \caption{\textbf{Spatially resolved CHSH measurements.} \textbf{(a)} Coincidence images recorded for two representative polarization projections in the benchmark setup. The left panel corresponds to the projection $(\bar{\beta'}=7\pi/2,\ \bar{\alpha}=\pi/2)$, while the right panel shows the projection $(\bar{\beta}=5\pi/8,\ \bar{\alpha'}=3\pi/4)$. \textbf{(b)} Spatial maps of the CHSH parameters $S_1$ (left) and $S_2$ (right), reconstructed from the corresponding sets of projective measurements. The white points mark regions where the calculated CHSH value exceeds $2\sqrt{2}$. These points arise from statistical fluctuations in areas with very low photon counts outside the main beam profile. \textbf{(c)} Regions of the transverse plane where the Bell inequality is violated ($|S|>2$), highlighted in yellow.}
    \label{fig:S3}
\end{figure*}

The experiment conducted with the PBOE elements is detailed in the main text. The benchmark measurement was performed by modifying the setup as illustrated in Fig.~\ref{fig:S2}. A D-shaped mirror placed in the crystal far-field was used to split the idler and signal into two separate paths, where projections on linearly polarized states were performed. The far-field was then imaged on the Tpx3Cam using a set of lenses and a PBS with half-wave plates (not shown) to have idler and signal beams on parallel paths. Fig.~\ref{fig:S3} show the results for the benchmark CHSH measurements.

\section{Slight off-focus placement of CHSH plate}
\begin{figure*}
\includegraphics[width=0.8\textwidth]{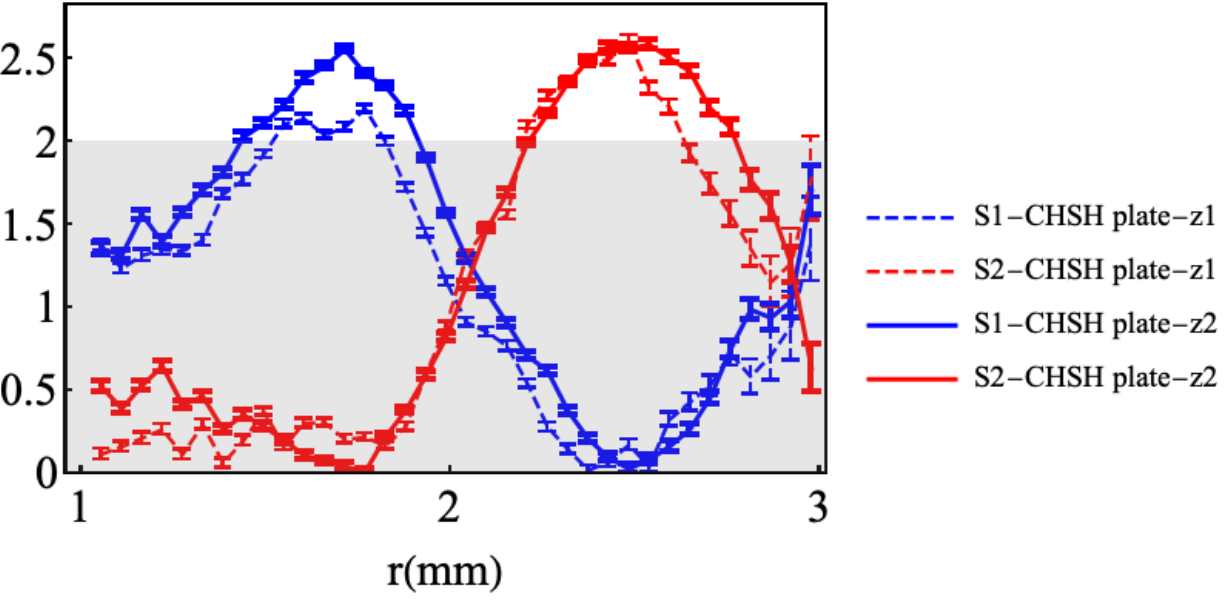}
\caption{Comparison of CHSH parameters $S1$ and $S2$, measured within a few millimeters along the propagation direction. The same data analysis procedure was applied to both datasets to assess the robustness of the technique to the CHSH plate position in the crystal’s farfield. A slight positional sensitivity is observed, with reduced CHSH violation for smaller radial values (closer to the centre) in the $z1$ measurements, with the plate placed at 2 mm away from focus.}
\label{fig:S4}
\end{figure*}

Fig.~\ref{fig:S4} shows a comparison of the results from the CHSH plate measurements where the plate is placed close to the far field. The two datasets were obtained by displacing the plate by $\sim 2$mm with respect to the longitudinal direction. 

\section{Programmable CHSH measurement using Spatial Light Modulator.}

\begin{figure*}
\includegraphics[width=0.8\textwidth]{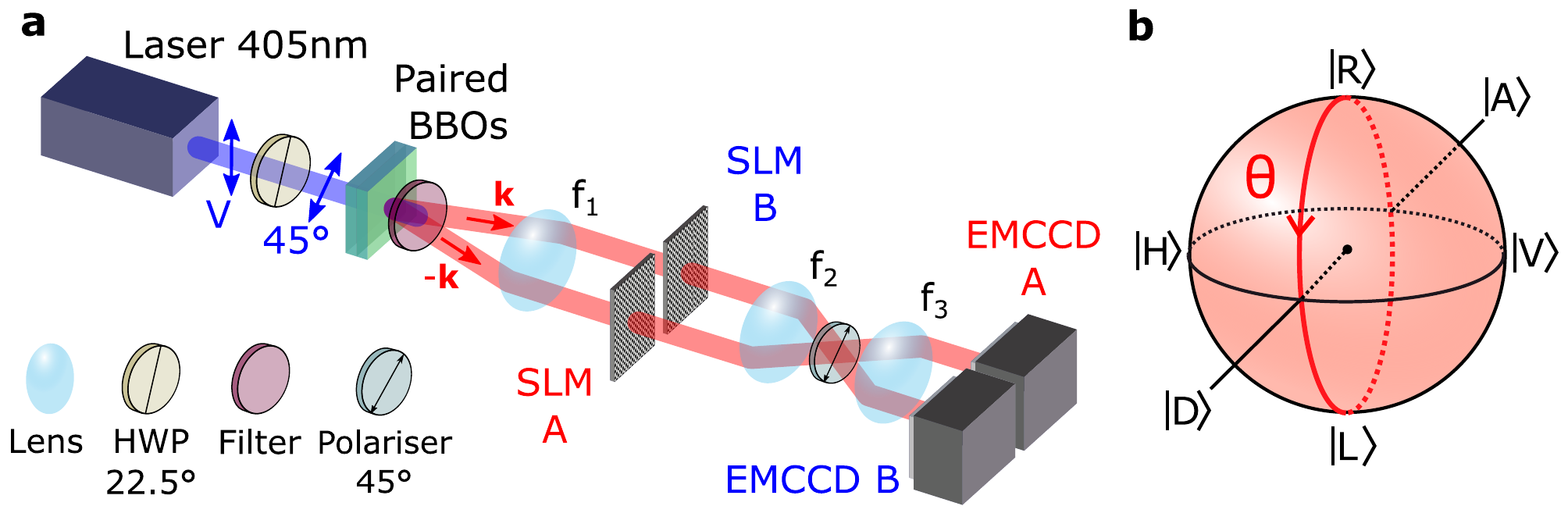}
\caption{\textbf{CHSH inequality using an SLM}. \textbf{a,} Light from a $405\,\mathrm{nm}$ laser diode, polarized at $45^\circ$, pumps a pair of orthogonally oriented $\beta$-barium borate (BBO) crystals ($0.5$\,mm thick each), generating spatially and polarization-entangled photon pairs via type-I SPDC. Pump photons are removed using long-pass and band-pass filters. The transverse momentum of the downconverted photons is imaged by lens $f_1$ onto an SLM (Holoeye Pluto-2-NIR-015), divided into two regions (SLM A and SLM B). Lenses $f_2$ and $f_3$ image the SLM plane onto an EMCCD camera, also split into two regions (EMCCD A and EMCCD B). A polarizer at $45^\circ$ is placed between lenses $f_4$ and $f_5$. Lens $f_1$ consists of three lenses ($45$\,mm, $125$\,mm, and $150$\,mm) arranged in a Fourier configuration, giving an effective focal length of $f_1=54$\,mm. The focal lengths of $f_2$ and $f_3$ are $150$\,mm and $100$\,mm, respectively. This configuration yields a direct mapping of $3\times3$ SLM pixels ($8 \times 8\,\mu m^2$ each) onto one camera pixel ($16 \times 16\,\mu m^2$). \textbf{b,} The red meridian on the Poincaré sphere represents the set of projective polarization measurements accessible at each camera pixel by tuning the phase $\theta$ applied to the corresponding SLM pixel.}
\label{fig:S5}
\end{figure*}

In this section, we propose an alternative experimental implementation of the setup shown in Fig.~\ref{fig:fig1}, using a spatial light modulator (SLM) instead of the CHSH plate. The corresponding experimental scheme is shown in Fig.~\ref{fig:S5}-\textbf{a}.
Here, the transverse momentum $\boldsymbol{k}$ of the photons is mapped onto the SLM pixels and subsequently re-imaged onto the camera ($3\times3$ SLM pixels correspond to $1$ camera pixel). Both the SLM and the camera are divided into two regions: photons with a negative transverse momentum component ($k_x < 0$) are shaped by SLM~A and detected by region~A of the camera, while photons with a positive transverse momentum component ($k_x > 0$) are shaped by SLM~B and detected by region~B of the camera. The SLM, based on parallel-aligned nematic liquid crystals, imparts a phase shift to the horizontal polarization component while leaving the vertical component unchanged. Because the SLM is followed by a $45^\circ$ polarizer, the phase shift applied at each SLM pixel defines a polarization projective measurement on photons of the associated transverse momentum $\boldsymbol{k}$, restricted to a meridian of the Poincaré sphere (Fig.~\ref{fig:S5}-\textbf{b}). In particular, this configuration enables projective measurements in the diagonal basis $\{\ket{D}, \ket{A}\}$ as well as in the circular basis $\{\ket{R}, \ket{L}\}$. Furthermore, the camera is an EMCCD, which enables single-photon detection and, by extension, the measurement of the spatially resolved second-order correlation function $G^{(2)}$~\cite{defienne_general_2018}. 

\begin{figure*}
\includegraphics[width=0.8\textwidth]{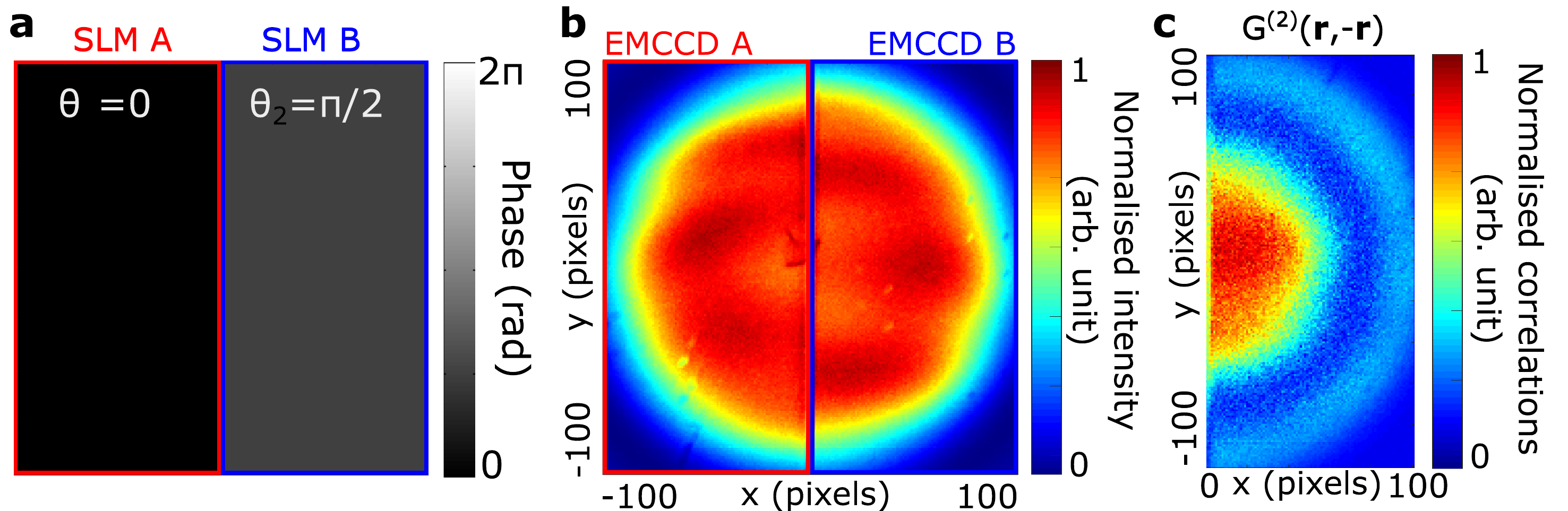}
\caption{\textbf{Example of a joint polarization measurement performed using the SLM.} 
\textbf{a,}~Two phase masks are displayed on distinct regions of the SLM, denoted SLM~A and SLM~B, which independently perform a projection polarization measurement on each photon of the pair ($x < 0$ for SLM~A and $x > 0$ for SLM~B). 
\textbf{b,}~Intensity image showing the SPDC emission pattern separated into two distinct regions, denoted EMCCD~A and EMCCD~B, corresponding to the associated SLM regions. 
\textbf{c,}~Correlation image $G^{(2)}(\boldsymbol{r}, -\boldsymbol{r})$ measured with the EMCCD, which, in this configuration, directly yields the result of the joint measurement $\bra{D} \rho \ket{R}$ for all pixel pairs.}
\label{fig:S6}
\end{figure*}

In the ideal case, the two-photon state at the SLM input plane, denoted $\ket{\Psi}$, is a state hyperentangled in both polarization and spatial degrees of freedom, which can be approximated as:
\[
\ket{\Psi} = \mathcal{N} \sum_{\boldsymbol{r} \in \mathcal{P}} \left [ \ket{HH}_{\boldsymbol{r},-\boldsymbol{r}} + e^{i\Psi(\boldsymbol{r})} \ket{VV}_{\boldsymbol{r},-\boldsymbol{r}} \right ]
\]
where $\mathcal{P}$ contains all output transverse momenta allowed by the phase-matching conditions, and $\Psi(\boldsymbol{r})$ represents a phase distortion inherent to the crystal geometry~\cite{hegazy_tunable_2015,hegazy_relative-phase_2017} (here, the spectral component is neglected, as the photons are assumed to be emitted at the same wavelength centered at $810$~nm). Focusing on polarization, the state generated by the crystal can be viewed as a polarization-entangled state where the two photons are distributed across numerous pairs of symmetric spatial modes $(\boldsymbol{r},-\boldsymbol{r})$. Consequently, by programming the SLM phase and measuring the anti-diagonal component of the correlation function, i.e. $G^{(2)}(-\boldsymbol{r}, \boldsymbol{r})$, with the EMCCD, we can perform parallel joint polarization measurements on the input state. For example, as shown in Fig.~\ref{fig:S6}-\textbf{a}, when half of the SLM is programmed to $\theta_1 = 0$ (corresponding to a $\ket{D}$ polarization measurement) and the other half to $\theta_2 = \pi/2$ (corresponding to a $\ket{R}$ polarization measurement), the correlation image $G^{(2)}(\boldsymbol{r}, -\boldsymbol{r})$ (Fig.~\ref{fig:S6}-\textbf{c}) directly yields the results of the joint measurements $\bra{D} \rho  \ket{R}$ for all pixel pairs (where $\rho$ is the density matrix associated to the unknown quantum state). In the following, we program the SLM phase to perform the $16$ joint measurements required to evaluate the CHSH criterion within a single correlation image.

\subsection{Preparation of the `Pizza' phase pattern}

To evaluate the CHSH inequality, we select two bases $\mathcal{A}=\{\ket{D},\ket{A}\}$ and $\mathcal{B}=\{\ket{R},\ket{L}\}$ for region A (SLM~A), and two rotated bases $\mathcal{A'}=\{\ket{D'},\ket{A'}\}$ and $\mathcal{B'}=\{\ket{R'},\ket{L'}\}$ for region B (SLM~B), defined as:
\begin{align}
    \ket{D} &= \frac{{1}}{\sqrt{2}} \left[ \ket{V} + \ket{H} \right]  \\
    \ket{A} &= \frac{{1}}{\sqrt{2}} \left[ \ket{V} - \ket{H} \right]  \\
    \ket{R} &= \frac{{1}}{\sqrt{2}} \left[ \ket{V} + i \ket{H} \right]  \\
    \ket{L} &= \frac{{1}}{\sqrt{2}} \left[ \ket{V} - i \ket{H} \right]  \\
    \ket{D'} &= \frac{\sqrt{3}}{\sqrt{2}} \left[ \ket{V} + \frac{1}{\sqrt{2}}\ket{H} \right]  \\
    \ket{A'} &= \frac{\sqrt{3}}{\sqrt{2}} \left[ \ket{V} - \frac{1}{\sqrt{2}}\ket{H} \right]  \\
    \ket{R'} &= \frac{\sqrt{3}}{\sqrt{2}} \left[ \ket{V} + \frac{i}{\sqrt{2}}\ket{H} \right]  \\
    \ket{L'} &= \frac{\sqrt{3}}{\sqrt{2}} \left[ \ket{V} - \frac{i}{\sqrt{2}}\ket{H} \right]  \\
\end{align}
The projective measurements corresponding to each basis vector are implemented by applying the following phase shifts to the SLM:
\begin{align}
    \ket{D} &\leftrightarrow 0 & \ket{D'} &\leftrightarrow \pi/4 \\
    \ket{A} &\leftrightarrow \pi & \ket{A'} &\leftrightarrow 5\pi/4 \\
    \ket{R} &\leftrightarrow \pi/2 & \ket{R'} &\leftrightarrow 3\pi/4 \\
    \ket{L} &\leftrightarrow 3\pi/2 & \ket{L'} &\leftrightarrow 7\pi/4 
\end{align}
To perform all $4\times4 = 16$ joint polarization measurements simultaneously, we construct a specific SLM phase mask termed the `Pizza' pattern (Fig.~\ref{fig:S7}). This pattern is centered at $\boldsymbol{r}=\boldsymbol{0}$ and divided into $32$ angular sectors (slices): 16 in region~A, corresponding to 4 different arrangements of the projective measurements $\{\ket{D},\ket{A}, \ket{R},\ket{L}\}$, and 16 in region~B, corresponding to 4 different arrangements of the projective measurements $\{\ket{D'},\ket{A'}, \ket{R'},\ket{L'}\}$. Ideally, each slice and its centrosymmetric counterpart (rotated by $180^\circ$) implement a joint projective polarization measurement between a vector from $\{\ket{D},\ket{A}, \ket{R},\ket{L}\}$ and a vector from $\{\ket{D'},\ket{A'}, \ket{R'},\ket{L'}\}$ for photon pairs located at $\boldsymbol{r}$ and $-\boldsymbol{r}$.

However, as shown in Fig.~\ref{fig:S6}-\textbf{c}, the projective measurement implemented by the SLM exhibits a radial dependence; specifically, the projection vector varies with $\boldsymbol{r}$. This variation arises from a phase aberration induced during the SPDC process, which manifests as a slight rotation of the measurement bases. This effect is significant because violating a CHSH-type inequality requires ensuring that correlation measurements are performed near the optimally chosen bases $\mathcal{A}, \mathcal{B}, \mathcal{A'}, \mathcal{B'}$. Any deviation from these bases could result in a failure to violate the inequality, thereby preventing conclusive verification of entanglement. To compensate for this distortion, we first characterize the phase distortion using the holography method described in Ref.~\cite{defienne2021polarization}.

\subsection{Phase distortion compensation}
To characterize the phase distortion $\Psi(\boldsymbol{r})$, we perform a quantum holographic measurement~\cite{defienne2021polarization}. To this end, a flat phase pattern $\theta_A = 0$ is displayed on SLM~A (Fig.~\ref{fig:S7}-\textbf{a}), while successive phase-shift patterns with values $0$, $\pi/2$, $\pi$, and $3\pi/2$ are displayed on SLM~B. Fig.~\ref{fig:S8}-\textbf{b} shows the correlation images $G^{(2)}(\boldsymbol{r}, -\boldsymbol{r})$ measured for each of the four configurations on SLM~B. The resulting reconstructed phase image corresponds exactly to the phase distortion $\Psi(\boldsymbol{r})$ (Fig.~\ref{fig:S7}-\textbf{c}).

\subsection{Bell inequality in a single image}
\begin{figure*}
\includegraphics[width=0.8\textwidth]{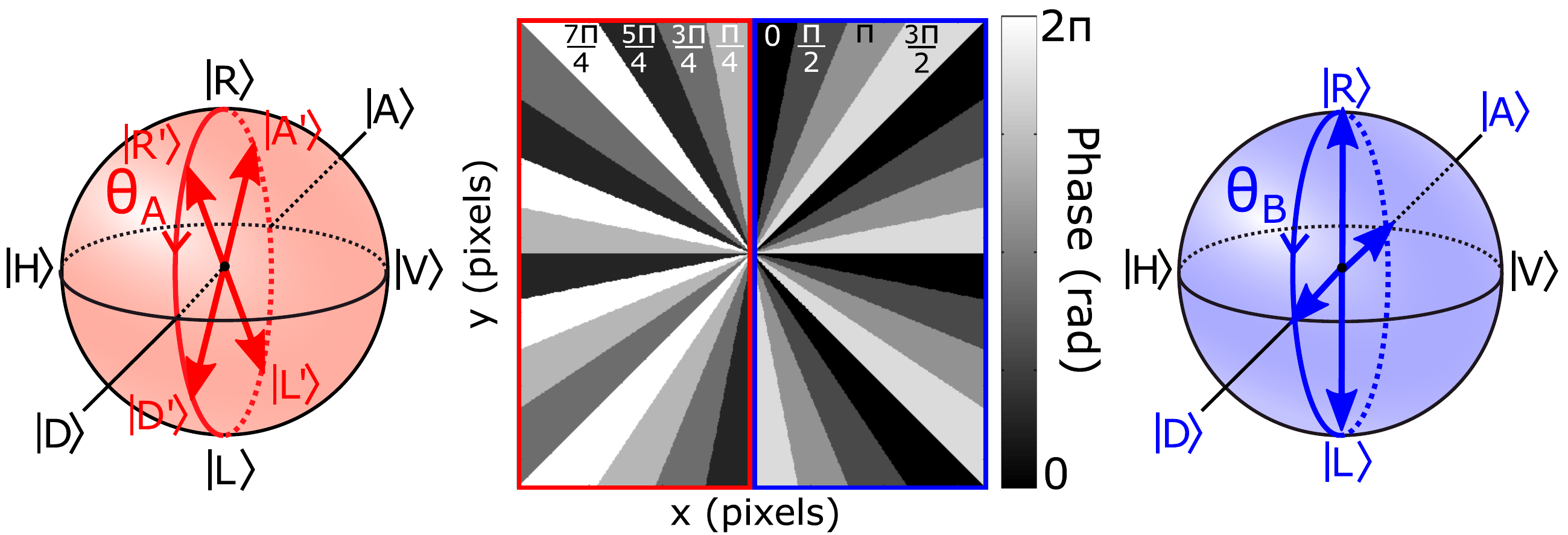}
\caption{\textbf{'Pizza' phase pattern for CHSH measurement.} An SLM pattern divided into 32 slices enables the parallel implementation of the 16 joint polarization measurements between the vectors of bases $\mathcal{A}=\{\ket{D},\ket{A}\}$ and $\mathcal{B}=\{\ket{R},\ket{L}\}$, and those of bases $\mathcal{A'}=\{\ket{D'},\ket{A'}\}$ and $\mathcal{B'}=\{\ket{R'},\ket{L'}\}$.}
\label{fig:S7}
\end{figure*}

\begin{figure*}
\includegraphics[width=0.8\textwidth]{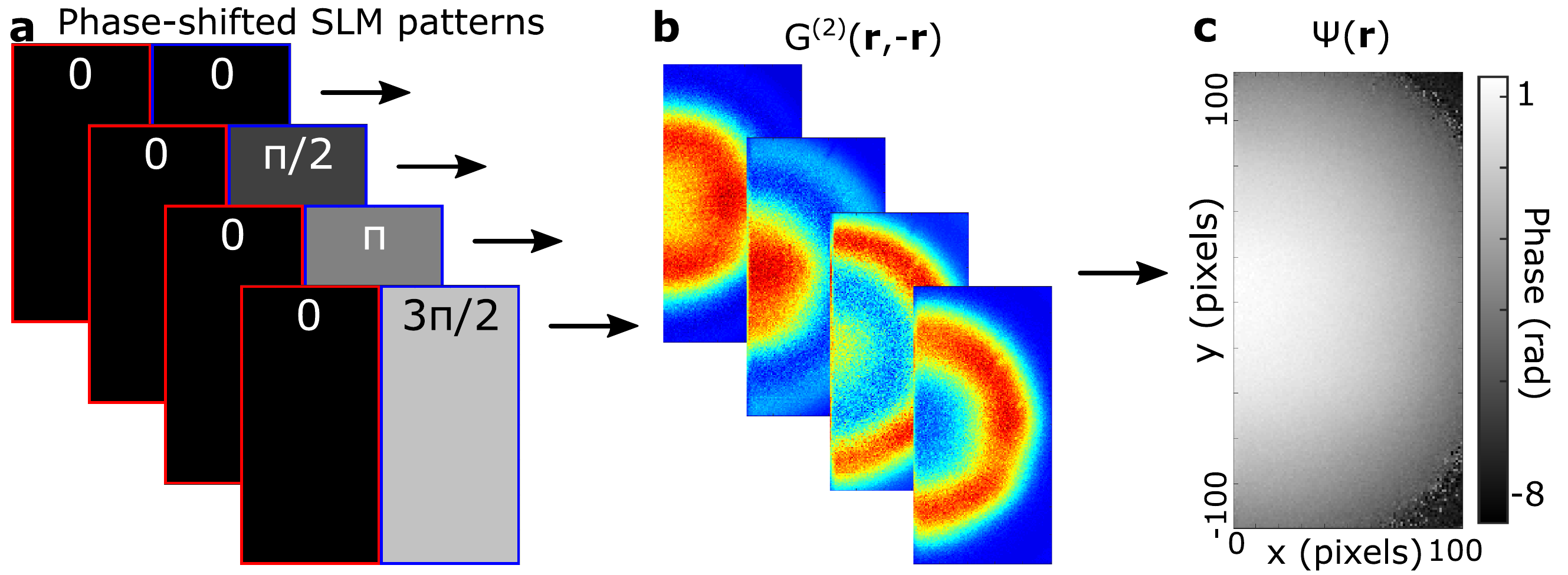}
\caption{\textbf{Phase distortion compensation.} Following the approach described in Ref.~\cite{defienne2021polarization}, uniform phase patterns with values of $0$, $\pi/2$, $\pi$, and $3\pi/2$ are programmed onto SLM~B (a), and the corresponding correlation images $G^{(2)}(\boldsymbol{r}, -\boldsymbol{r})$ are measured. These correlation images are subsequently recombined using a phase-shifting holography formula to reconstruct the phase distortion $\Psi(\boldsymbol{r})$ affecting the projective measurements performed by the SLM.}
\label{fig:S8}
\end{figure*}

As shown in Fig.~\ref{fig:fig4}-\textbf{a}, the phase distortion image $\Psi(\boldsymbol{r})$ is fitted by a quadratic function of the form $\Psi(x, y) = 4.69k_x^2 + 5.04k_y^2 + 0.02$. This profile is then superimposed onto the `Pizza' SLM pattern to correct the measurement bases. Once the pattern is programmed on the SLM, diffraction effects at the boundaries of the slices become visible in the intensity image (Fig.~\ref{fig:fig4}-\textbf{b}). We then measure the correlation image $\Gamma(\boldsymbol{r}, -\boldsymbol{r})$ (Fig.~\ref{fig:S7}-\textbf{c}), where we observe a reduced radial dependence, i.e. the values within each slice appear more homogeneous along the radial direction.

Fig.~\ref{fig:fig4}-\textbf{c} contains the $16$ joint polarization measurement values required to compute the CHSH $S$-parameter. To avoid artefacts arising from diffraction at the boundaries, we calculate the mean values and standard deviations by selecting positions $\boldsymbol{r}$ located strictly within the interior of each slice, as detailed in Fig.~\ref{fig:fig4}-\textbf{c}. Table~\ref{tab:chsh_values} presents the mean values and standard deviations associated with each selected region. From these joint measurements, we calculate the correlation coefficients between the different basis pairs using the formula:
\begin{equation}
    E(\mathcal{A},\mathcal{A'}) = \frac{\bra{D}\rho \ket{D'}-\bra{D}\rho \ket{A'}-\bra{A}\rho \ket{D'}+\bra{A}\rho \ket{A'}}{\bra{D}\rho \ket{D'}+\bra{D}\rho \ket{A'}+\bra{A}\rho \ket{D'}+\bra{A}\rho \ket{A'}},
\end{equation}
where $\bra{D}\rho \ket{D'}$ denotes the joint measurement outcome obtained from the state $\rho$ between mode $\ket{D}$ of basis $\mathcal{A}$ and mode $\ket{D'}$ of basis $\mathcal{A'}$. This formula generalizes to the three other correlation coefficients between the bases: $E(\mathcal{B},\mathcal{A'})$, $E(\mathcal{A},\mathcal{B'})$, and $E(\mathcal{B},\mathcal{B'})$. The values of the different coefficients and their standard deviations are reported in Table~\ref{tab:chsh_values}. Finally, the parameter $S$ is calculated using the formula:
\begin{equation}
    S = |E(\mathcal{A},\mathcal{A'}) - E(\mathcal{A},\mathcal{B'})| + |E(\mathcal{B},\mathcal{A'}) + E(\mathcal{B},\mathcal{B'})|.
\end{equation}
We observe a direct violation of the CHSH inequality with $S = 2.5586 \pm 0.1645$~\cite{clauser_proposed_1969}.

\begin{table}[htbp]
\centering
\caption{\textbf{Joint polarization measurements and basis correlation coefficients.} }
\label{tab:chsh_values}

\begin{tabular}{l cc @{\hspace{1.5em}} cc}
\toprule
\multicolumn{5}{c}{\textbf{Joint polarization measurements (arbitrary units)}} \\
\midrule
 & $\ket{D'}$ & $\ket{A'}$ & $\ket{R'}$ & $\ket{L'}$ \\
\midrule
$\ket{D}$ & $77.9660 \pm 19.7407$ & $269.1489 \pm 27.9431$ & $314.9419 \pm 31.3285$ & $28.9057 \pm 22.0664$ \\
$\ket{A}$ & $259.3669 \pm 22.2723$ & $75.2745 \pm 19.4538$ & $31.6214 \pm 19.7320$ & $307.6156 \pm 30.0146$ \\
\addlinespace[1.2ex] 
$\ket{R}$ & $299.1731 \pm 29.0154$ & $51.9013 \pm 20.9962$ & $246.3462 \pm 22.9330$ & $75.7213 \pm 18.1522$ \\
$\ket{L}$ & $61.0554 \pm 28.2897$ & $302.3889 \pm 22.6504$ & $88.2140 \pm 21.5380$ & $247.5014 \pm 22.5245$ \\

\midrule
\multicolumn{5}{c}{\textbf{Basis correlation coefficients}} \\
\midrule
\multicolumn{2}{r}{$E(\mathcal{A}, \mathcal{A'}) = -0.5505 \pm 0.0757$} & & \multicolumn{2}{l}{$E(\mathcal{A}, \mathcal{B'}) = 0.8228 \pm 0.0996$} \\
\multicolumn{2}{r}{$E(\mathcal{B}, \mathcal{A'}) = 0.6838 \pm 0.0783$} & & \multicolumn{2}{l}{$E(\mathcal{B}, \mathcal{B'}) = 0.5016 \pm 0.0727$} \\
\bottomrule
\end{tabular}
\end{table}

\end{document}